\documentclass[article,showpacs]{revtex4}
\usepackage{amssymb}
\usepackage{amsfonts}
\usepackage{amsmath}
\usepackage{graphicx}
\usepackage{epsfig}
\usepackage{dcolumn}
\usepackage{bm}
\usepackage{dcolumn}%
\begin{document}
\title{ How HF solutions for the TB interacting electrons in 2D  predict SCES properties
and  suggest phenomenology  for attaining RTS }
\author{A. Cabo Montes de Oca$^{*,**}$, N. H. March$^{*,***,****}$ and A.
Cabo-Bizet$^{*,*****}$ \bigskip}
\affiliation{$^{*}$International Centre for Theoretical Physics, Strada Costiera 11,
Miramare, Trieste, Italy}
\affiliation{$^{**}$ Departamento de F\'isica Te\'orica, Instituto de Cibern\'etica
Matem\'atematica y F\'{\i}sica (ICIMAF), Calle E, No. 309, entre 13 y 15,
Vedado, La Habana, Cuba}
\affiliation{$^{***}$ Department of Physics, University of Antwerp, Antwerp, Belgium }
\affiliation{$^{****}$ Oxford University, Oxford, England. }
\affiliation{$^{*****}$ Centro de Estudios Aplicados al Desarrollo Nuclear, Calle 30, esq a
5ta Avenida, Miramar, La Habana, Cuba}

\begin{abstract}
\noindent Former results for a Tight-Binding (TB) model of CuO planes in
$La_{2}CuO_{4}$ are reinterpreted here to underline their wider implications.
It is noted that physical systems being appropriately described by the TB
model can exhibit the main strongly correlated electron system
(SCES) properties, when they are solved in the HF approximation, by also
allowing crystal symmetry breaking effects and non-collinear spin orientations
of the HF orbitals. It is argued how a simple 2D square lattice system of
Coulomb interacting electrons can exhibit insulator gaps and pseudogap states,
and quantum phase transitions as illustrated by the mentioned former works. A
discussion is also presented here indicating the possibility of attaining room
temperature superconductivity, by means of a surface coating with water
molecules of cleaved planes of graphite, being orthogonal to its c-axis. The possibility
that  2D arrays of quantum dots can give rise to the same effect is also proposed
to consideration.  The analysis also furnishes  theoretical insight to solve the Mott-Slater debate, at
least for the $La_{2}CuO_4$ and TMO band structures. The idea is to apply a
properly non-collinear GW scheme to the electronic structure calculation of
these materials. The fact is that the GW approach can be viewed as a HF
procedure in which the screening polarization is also determined. This  directly
indicates the possibility of predicting the assumed dielectric constant in the previous works.
Thus, the results seem to identify that the main correlation properties in
these materials are determined by screening. Finally, the conclusions also seem to
be of help for the description of the experimental observations of
metal-insulator transitions and Mott properties in atoms trapped in planar
photonic lattices.

\end{abstract}

\maketitle


\section{Introduction}

Band structure calculation is a central part of modern Solid State theory. The
area has a long history in the literature along the continuous quest about the
structure of matter \cite{mott1,slater1,slater,
bmuller,dagoto,almasan,yanase,vanharlingen,damascelli,pickett,
Burns,imada,freltoft,deBoer,peierls,anderson1,hubbard,adersson,fradkin,
gutzwiller,gutzwiller1,rice,kohn,kohn1,terakura,
singh,szabo,fetter,matheiss,mott}. A central open problem in this field is the
connection between the so called $first$ $principles$ (or $ab-initio$) schemes
with the Mott phenomenological methods \cite{mott,slater,imada,yanase,dagoto}.
Materials which had taken part in the center of the debate between the two
conceptions were the transition metal oxides (TMO) and the HTSC materials
\cite{pickett,almasan,
imada,yanase,anderson1,terakura,hubbard,gutzwiller,gutzwiller1}. One
particular substance in this set is the early superconducting material
$La_{2}CuO_{4}$. Its first band structure calculations predicted a metal and
paramagnetic characters, which were drastically different from its
experimentally known insulating and antiferromagnetic nature \cite{matheiss}.
These two qualities are strong correlation properties which can't be derived
from an $ab-initio$ HF independent particle description
\cite{anderson1,terakura,augustoref}. \ Seemingly paradoxical, in Refs.
\cite{pla,symmetry,last} a single band Hartree-Fock (HF) study was able to
produce these strong correlation properties of $La_{2}CuO_{4}$, as independent
particle ones, arising from a combination of crystal symmetry breaking with an
$entangled$ spin-spatial structure of the single particle states. \ However,
the conclusion is not strange, after taking into account that the procedure
employed was not an $ab-initio$ method. That is: the mean field approach was
applied to a model in which the Coulomb interaction was screened by means of a
phenomenological dielectric constant in order to match the band width of the
spin polarized solution with the one associated to the single band crossing
the Fermi energy in the Matheiss band structure evaluations for $La_{2}%
CuO_{4}.$ \ Therefore the use of the HF scheme in these works has the same
degree of applicability as the similar use of the mean field method for the
evaluation of the AF order and the quasiparticles in the Hubbard models, which
are considered to embody the SCES properties of the Mott substances, such as
the $La_{2}CuO_{4}$ system.\ The direct inclusion of correlation effects in
the considered model is clearly represented by the phenomenological way of
including screening effects, which clearly add correlation to the model. An
important difference with the Hubbard models is however, the fact that by not
adopting the Hubbard strong approximations, the scheme was afterwards able to
predict the existence of the pseudogap state (\cite{pla,symmetry,last}), the
nature of which is today still strongly debated \cite{laughlin}. \ We consider
that a special merit of the present approach is that it identifies the
relevant role of a particular form of correlations: namely screening, in
determining the emergence of the insulating and AF character of the TMO and
HTSC materials. \ The discussion in Ref. \cite{pla,symmetry,last} considered a
single band model of $La_{2}CuO_{4}$. Then, the derivation of the same results
from an $ab-initio$ band structure calculation could represent an appreciable
contribution to clarify the links between the Mott and the $ab-initio$
descriptions. \ This problem will be considered further elsewhere.

The present work is devoted to evidence general implications of the system investigated in Ref. \cite{pla,symmetry,last}: the Tight-Binding model
of 2D interacting electrons. \ The wide range of applicability of this model
in 2D systems suggests that the appearance of SCES properties already in the
HF approximation, in the context of the model as controlled by the overlapping
and other parameters, can be of help in the search for physical systems of
interest in nanotechnology. Then, after shortly reviewing Dirac's fully
unconstrained formulation of the HF problem, the mean field equations for the
square 2D lattice model employed in \cite{pla,symmetry,last} are  written
by allowing more freedom in the values of the parameters. Another main element
is to allow a weaker translation symmetry. That is, the imposition of Bloch
conditions on the HF self-energy modes, but in smaller sublattices being
fractions of the total. We present the solutions obtained for the band
structures of the TB model at half filling when the overlapping constant is
varied by changing the width of the employed TB Gaussian orbitals. At small
overlapping values the system developed an insulator gap which afterwards
tends to close when the overlapping increases. This suggests the presence of
an insulator to metal transition at even larger overlapping which however, lies outside
the scope of the TB approximation. Apart from the parameter controlling
overlapping, the others were chosen at the same values as in the model
discussed in Ref. \cite{pla,symmetry,last}. It can be noted that in addition
with the connection with htsc cuprates, from which the present analysis
emerged, the received conclusions seem to furnish a clear understanding for
the experimental observations of metal-insulator transitions and Mott
properties in systems of atoms trapped in planar photon lattices
\cite{opt1,opt2,opt3}.

The second issue examined consists in the application of the ideas in the
model to develope an explanation of recent experimental results indicating the
existence of granular room temperature superconductivity (RTS) as induced by chemisorption of water
molecules on graphite powder \cite{RT}. These experiments detect the
hysteresis loops shown by suspensions of the powder, which indicate the
presence of magnetic properties in the grains. Due the presence of Oxygen in the
water, one possibility could be to imagine the role of the presence of
$O_{2}$ molecules in the system, due to the known ferromagnetic moment in
this compound (See Ref. \cite{march}). However, it is also possible that the
powder becomes superconducting after doped with water \cite{RT}. We will
explore here this last possibility, as motivated by observing some features
with seem to link the system with the TB model of interacting electrons.

Recent $ab-initio$ evaluations of the deposition of water on graphene had been
performed in Refs. \cite{antwerp,wehling}. The authors of that work remark that the
deposition of a single layer of water on the graphene leads to a 2D planar
hexagonal crystal of water molecules sitting nearly 3.4 $A^{0}$ above the
graphene. \ The centers of the Oxygen atoms stay directly over the centers of
the hexagons formed by the carbon atoms. They explain that the periodic
potential created by the water molecules on graphene does not cause it to
become metallic. \ However, they also argue that if a substrate of $SiO_{2}$
is added below the graphene, then the system might become conducting. \ This
is a point that we consider as affording an opportunity for a TB model to be
of relevance in describing the system. The idea is that graphite with a
monolayer of water over the last graphene layer in it, can be considered as a
graphene layer with a substrate of the same graphite. \ But as the simulations
indicated, a substrate can help the last layer to become metallic. This view
is also supported by the fact that the same graphite is a semi-metal. Therefore,
assuming that the same predictions of SCES\ can be obtained from a TB model of
2D electrons with a lattice showing hexagonal structure, the last hexagonal
graphene layer on the surface of graphite (which is separated by a large
distance from the other layers) could be eventually seen as TB metal upon
which the deposited 2D crystal of water molecules exerts a strong dipolar
periodic potential. \ As a consequence, it could be expected that this field
may lead to create a TB binding band eventually producing an insulator and its
corresponding pseudogap state along the lines discussed here.   The elements
 of the discussion of the water doped graphite system also suggested the possibility
that  2D arrays of quantum dots can give rise to the same effect, if no strong
 experimental limitations exist on the optimization of the ratio between the
squared hopping parameter for the lattice of QDots and the width of the TB band
determined  by  the potential generated by them.

The presentation proceeds as follows. The section 2 reviews the fully
unrestricted HF procedure. Further, Section 3 presents the results of the insulator band
structure  with varying overlapping in the square
two-dimensional TB model with interacting electrons. Next,  section 4 discusses the
possibility of obtaining room temperature superconductivity in water doped
graphite surfaces and 2D lattice arrays of quantum dots. Finally, the results are summarized.

\section{ The fully unrestricted Dirac's HF scheme}

Let us briefly review here the fully unrestricted HF procedure employed in the
following discussion. The state of a system of N particles is described by the
wavefunction depending on the particles spinor and spatial coordinates
$f_{n}(x_{1};s_{1},...,x_{N};s_{N})$, where $n$ represents the set of quantum
numbers indexing the state \cite{fetter}. The HF approximation is defined by
assuming the above mentioned state as expressed by a linear combination of
antisymmetric products of N orthonormalized orbitals $\phi_{k_{i}}(x_{i}%
,s_{i}),$ with $i=1,...,N$. Each one of the orbitals defines a single particle
state. As usual, in what follows the word coordinate will mean the spatial as
well as the spinor ones \cite{fetter,slater,szabo}.

Let
\begin{align}
\label{hamiltoniano}\hat{h}(x_{1},...x_{N})=\sum_{i}\hat{h}_{0}(x_{i}%
)+\frac{1}{2}\sum_{j\neq i}V(x_{i},x_{j}),
\end{align}
be the 2D electron system Hamiltonian, including the kinetic term, plus the interaction
with the environment Hamiltonian $\hat{h}_{0}$ and the Coulomb interaction potential $V$
among pairs of electrons. The HF equations of motion for the system follow
after imposing the extremum condition on the mean energy functional, in the
form
\begin{align}
\lbrack\ \hat{h}_{0}(x)+\sum_{\eta_{1}}\sum_{s^{\prime}}\int d^{2}x^{\prime
}\phi_{\eta_{1}}^{\ast}(x^{\prime},s^{\prime})V(x,x^{\prime})\phi_{\eta_{1}%
}(x^{\prime},s^{\prime})\ ]\ \phi_{\eta}(x,s)  & \label{HF}\\
-\sum_{\eta_{1}}[\ \sum_{s^{\prime}}\int d^{2}x^{\prime}\phi_{\eta_{1}}^{\ast
}(x^{\prime},s^{\prime})V(x,x^{\prime})\phi_{\eta}(x^{\prime},s^{\prime
})\ ]\phi_{\eta_{1}}(x,s)  & =\varepsilon_{\eta}\ \phi_{\eta}%
(x,s),\ \ \ \ \ \ \nonumber
\end{align}
where $\eta=k_{1},...,k_{N}$ is the label in the basis of solutions. As usual the
self-consistent Hamiltonian has two components: the direct and the exchange potentials
 \cite{dirac}. The HF energy of the $N$ electron system
and the interaction energy of an electron in the $\eta$ state with the
remaining ones, are given by the expressions%

\begin{align}
E_{HF}  & =\sum_{\eta}\langle\eta|\hat{h}_{0}|\eta\rangle+\frac{1}{2}%
\sum_{\eta,\eta_{1}}\langle\eta,\eta_{1}|V|\eta_{1},\eta\rangle-\frac{1}%
{2}\sum_{\eta,\eta_{1}}\langle\eta,\eta_{1}|V|\eta,\eta_{1}\rangle
,\label{enerpar}\\
a_{\eta}  & =\frac{1}{2}\sum_{\eta_{1}}\langle\eta,\eta_{1}|V|\eta_{1}%
,\eta\rangle-\frac{1}{2}\sum_{\eta_{1}}\langle\eta,\eta_{1}|V|\eta,\eta
_{1}\rangle.\nonumber
\end{align}

The bracket terms in (\ref{enerpar}) represent the following integrals:
\begin{align}
\langle m|\hat{h}_{0}|p\rangle & \equiv\sum_{s}\int d^{2}x\ \phi_{m}^{\ast
}(x,s)\ \hat{h}_{0}(x)\ \phi_{p}(x,s),\\
\langle m,n|V|o,p\rangle & \equiv\sum_{s,s^{\prime}}\int d^{2}xd^{2}x^{\prime
}\ \phi_{m}^{\ast}(x,s)\phi_{n}^{\ast}(x^{\prime},s^{\prime})\ V(x,x^{\prime
})\ \phi_{o}(x^{\prime},s^{\prime})\phi_{p}(x,s),\nonumber
\end{align}
where $m,n,o$ and $p$ denote any one of the possible quantum number indices.

It can be noted that the system of equations (\ref{HF}) shows no constraint
whatsoever on the spin and orbital structure of the single particle HF states,
because it is written without imposing a spatially absolute direction for the
spin quantization of the single electron orbitals. This fully unrestricted
invariant formulation of the self-consistent HF procedure was first introduced
by Dirac in Ref. \cite{dirac}.

In general, finding the solution of (\ref{HF}) is a complicated task. The
iterative method is one of the most frequently employed for solving this kind
of system and it is usually complemented by the imposition of symmetry
restrictions that reduce the space of states to be sampled. However, such
constraints can avoid obtaining special solutions not satisfying the fixed
 conditions. Although in some cases the constraints can be satisfied
by the minimal energy state, the procedure can hide the existence of
interesting excited states and inclusive could wrongly predict the excitation
features in some cases, as had been seen in Ref. \cite{pla,symmetry,last} . A
very common example of such restrictions, which are usually employed in band
theory and quantum chemistry calculations, is to consider that single particle
solutions of (\ref{HF}) have the spin quantized in a fixed direction at every
point of the space \cite{matheiss,szabo}. That means, they satisfy
\[
\phi_{\mathbf{k}}(x,s)=%
\begin{cases}
\phi_{\mathbf{k}}^{\alpha}(x)\ u_{\uparrow}(s) & \text{$\alpha$ state},\\
\phi_{\mathbf{k}}^{\beta}(x)\ u_{\downarrow}(s) & \text{$\beta$ state},
\end{cases}
\]
where $u_{\uparrow\downarrow}$ represent the Pauli spinors with spin up and
down in a certain globally fixed direction, respectively. If the spacial
functions $\phi_{\mathbf{k}}^{\alpha}$ and $\phi_{\mathbf{k}}^{\beta}$ are the
same, the HF calculation is called a "restricted" one, and if they are
different, the procedure is called "unrestricted" \cite{szabo}. It was thus of
interest to examine the consequence of considering the possible existence of
non-separable single particle states being solutions of the HF problem in the
context of a system showing antiferromagnetism. This was the main motivation
in starting the investigation leading to the works in Refs.
\cite{pla,symmetry,last}. \ Afterwards, the insulator gap appeared as a surprise.

Another important type of constraints posed on $ab-initio$ band theory
evaluations is the a priori impositions of crystal symmetries. To beforehand
impose a symmetry on the assumed to be searched solution, although it could be
suggested in the studied system, has the risk of hiding a possible spontaneous
breaking of that invariance. This could occur due to the reduction of the
space of orbitals in which the  search is done. In that case it can turn out that
the obtained solutions will not be an absolute extremal of the energy
functional, but a conditional one due to the fixed symmetry constraint.

\section{The 2D square TB model}

\begin{figure}[h]
\vspace{.1cm}
\begin{center}
\includegraphics[width=1.3in,height=1.3in]{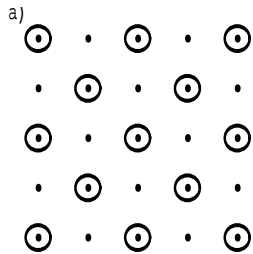}
\end{center}
\caption{ The figure illustrates the decomposition of the full lattice in two sublattices
 in order to allow a lattice translation symmetry breaking.}
\label{bravais}%
\vspace{.1cm}
\end{figure}

The free hamiltonian of the model is considered in the form
\begin{align}
\hat{h}_{0}(\mathbf{x})  & =\frac{\hat{\mathbf{p}}^{2}}{2m}+W_{\gamma
}(\mathbf{x})+F_{b}(\mathbf{x}),\label{freehamiltonian}\\
W_{\gamma}(\mathbf{x})  & =W_{\gamma}(\mathbf{x}+\mathbf{R}),\nonumber\\
F_{b}(\mathbf{x})  & =\frac{e^{2}}{4\pi\epsilon\epsilon_{0}}\sum_{\mathbf{R}%
}\int d^{2}y\frac{\rho_{b}(\mathbf{y}-\mathbf{R})}{|\mathbf{x}-\mathbf{y}%
|},\ b\ll p,\nonumber
\end{align}
where $\hat{\mathbf{p}}^{2}$ is the particle's squared momentum operator;
\emph{m} is the particle mass; $\epsilon_{0}$ is the vacuum permittivity and
the square lattice to which the free electrons bind is defined by
\[
\mathbf{R}=%
\begin{cases}
n_{x_{1}}p\ \hat{\mathbf{e}}_{x_{1}}+n_{x_{2}}p\ \hat{\mathbf{e}}_{x_{2}},\\
\text{ with }n_{x_{1}}\text{ and }n_{x_{2}}\in\mathbb{Z}.
\end{cases}
\]
The unit vectors $\hat{\mathbf{e}}_{x_{1}}$ and $\hat{\mathbf{e}}_{x_{2}}$
point along the directions defined by the lattice's nearest neighbors, see
figure \ref{bravais}. The lattice parameter $p$ is supposed now to be free,
at variance with the case in Refs. \cite{pla,symmetry,last}, where it was fixed to
the known distance between Cu nearest neighbors in CuO planes of $La_{2}%
CuO_4:$ $p\ \approx$ 3.8 \AA \cite{pickett}.

The term F$_{b}$ represents the interaction between the particles and the
\textquotedblright jellium\textquotedblright\ neutralizing charges. The
jellium is modeled as a gaussian distribution of positive charges
\[
\rho_{b}(\mathbf{y})=\frac{1}{\pi b^{2}}\exp(-\frac{\mathbf{y}^{2}}{b^{2}}),
\]
sitting at each point of the square lattice and $b$ defines the degree of
spreading around each of the lattice points. \ We also include the Coulomb
interaction among pairs of particles in the form
\begin{equation}
V(\mathbf{x},\mathbf{y})=\frac{e^{2}}{4\pi\epsilon\epsilon_{0}}\frac
{1}{|\mathbf{x}-\mathbf{y}|},\label{coulombio}%
\end{equation}
which, as remarked above, takes into account a dielectric constant $\epsilon$
that is considered as generated by the effective environment.

The full square lattice will be divided in the two sublattices shown in figure
\ref{bravais}, and the single particle eigenstates will be constrained to be
eigenfunctions only of the reduced group of lattice translations
transforming each of those sublattices on itself. \ This represent only a limited
reduction of translation symmetry breaking in the final solution, since still the
HF orbitals will be forced to be Bloch functions in the sublattices. However,
the freedom allowed is sufficient to permit the appearance of
antiferromagnetic solutions. We also will limit the
discussion here to these important cases, in order to maintain the simplicity of
the discussion in \cite{pla,symmetry,last}.\ The points of the two sublattices
will be indexed as r = 1, 2, and are defined as follows
\begin{align}
\mathbf{R}^{(r)}  & =\sqrt{2}n_{1}p\ \hat{\mathbf{q}}_{1}+\sqrt{2}n_{2}%
p\ \hat{\mathbf{q}}_{2}+\mathbf{q}^{(r)}\label{e:subred}\\
& \ \ \ \ \ \ \text{ with }n_{1}\text{ and }n_{2}\ \in\mathbb{Z},\nonumber\\
\mathbf{q}^{(r)}  & =%
\begin{cases}
\mathbf{0}, & \text{ if\ r=1},\\
p\ \hat{\mathbf{e}}_{x_{1}}, & \text{ if r=2},
\end{cases}
\nonumber
\end{align}
in which $\hat{\mathbf{q}}_{1}$ and $\hat{\mathbf{q}}_{2}$ are the basis
vectors on each one of them.

The condition of being a Bloch function in the sublattices is represented by
the following eigenvalue equations of the operators $\hat{T}_{\mathbf{R}%
^{(0)}}$ belonging to the reduced discrete translation group which transforms
a given sublattice on itself:
\[
\hat{T}_{\mathbf{R}^{(0)}}\phi_{\mathbf{k},l}=\exp(i\ \mathbf{k}%
\cdot\mathbf{R}^{(0)})\phi_{\mathbf{k},l}.
\]

\begin{figure}[h]
\begin{center}
\includegraphics[width=1.6in,height=1.6in]{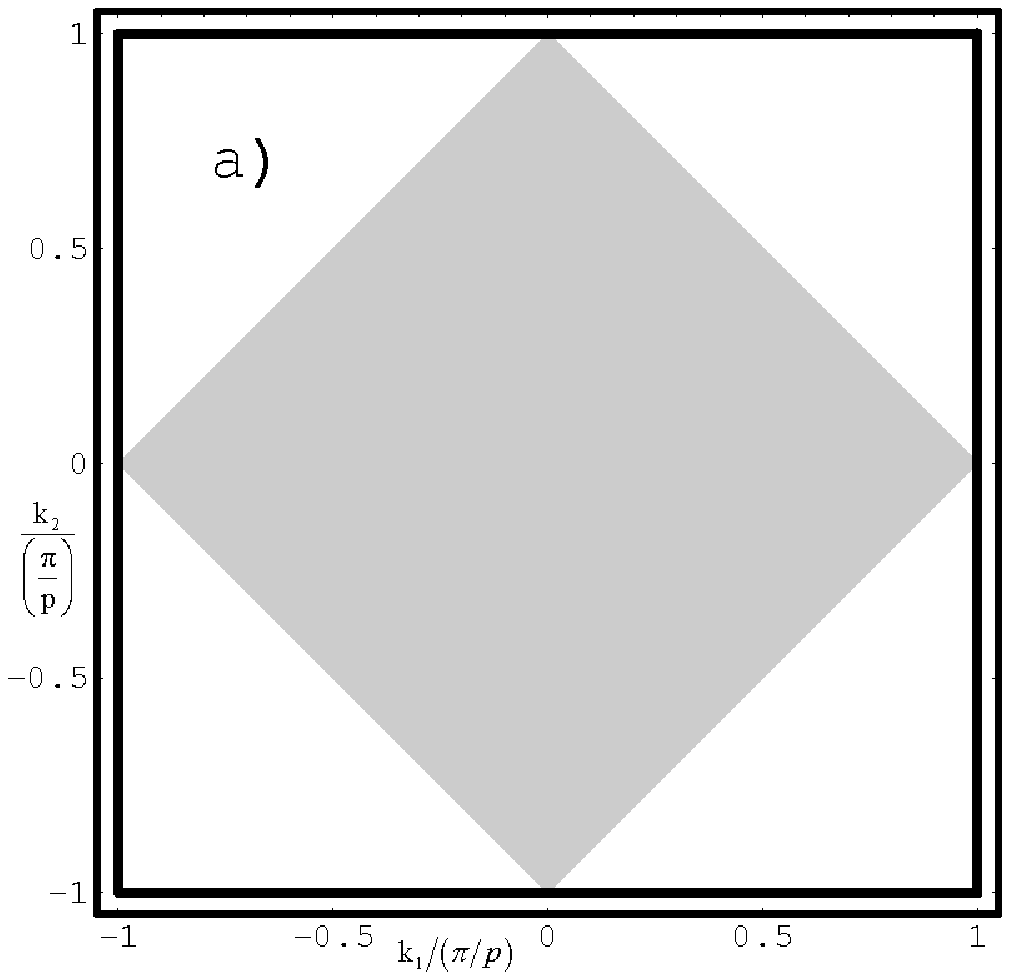}
\includegraphics[width=1.6in,height=1.6in]{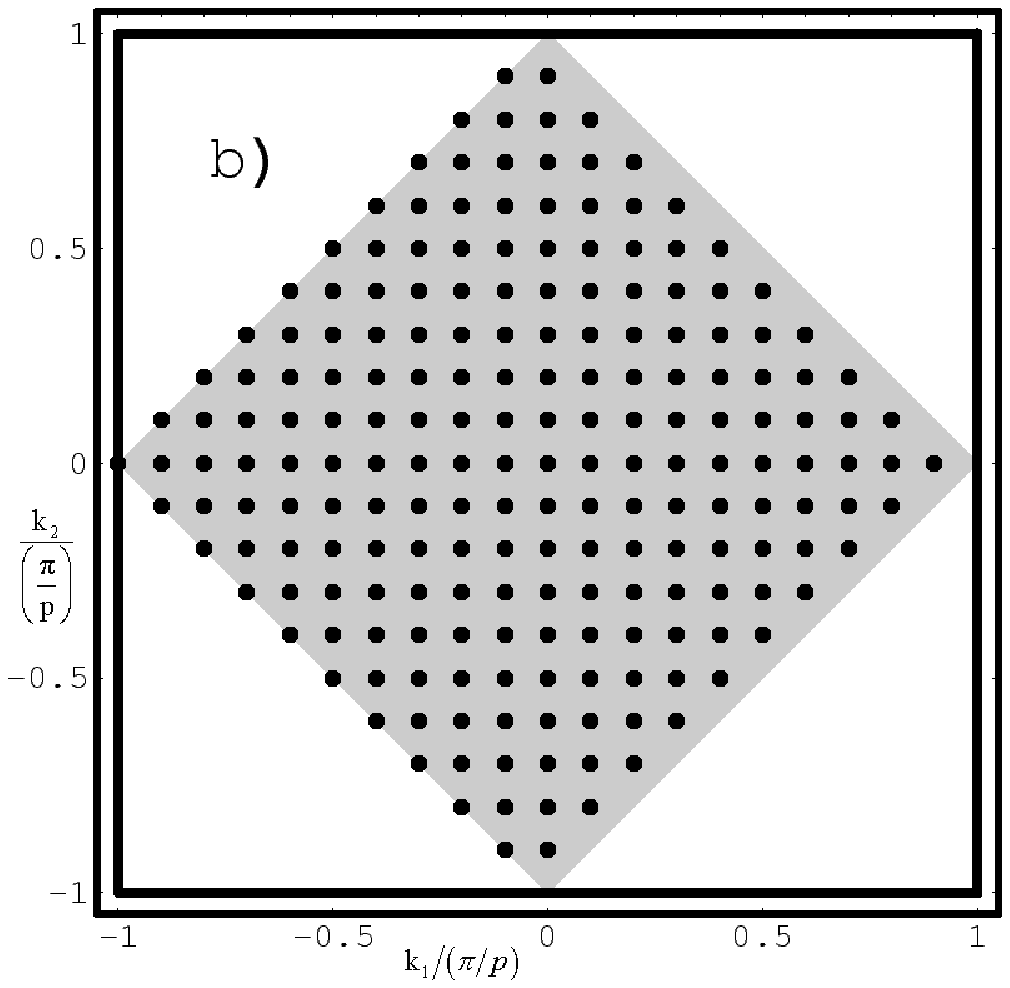}
\end{center}
\caption{Figure a) shows the continuous zone of momenta satisfying the periodic boundary conditions
 for the infinite lattice. Figure b) illustrates the discrete set of momenta which obey the imposed
 periodic boundary conditions
 in a finite square set of lattice points.
 }
\label{f:ZB}%
\end{figure}

For the Bravais lattice of the infinite crystal the Brillouin zone (B.Z.)
associated to $\hat{T}_{\mathbf{R}^{(0)}}$ becomes the shadowed region in
figure \ref{f:ZB} a). The whole square in this figure represents the B.Z.
associated to the group of translations leaving invariant the whole square
lattice. However, in order to practically solve the equations it should be
imposed periodic boundary conditions on the $\phi_{\mathbf{k},l}$ in the
lattice boundaries $x_{1}=-Lp, Lp$ \ and $\ x_{2}=-Lp, Lp$ (see figure
\ref{f:ZB} b)). This condition determines the allowed set of momenta
$\mathbf{k}$
\[
\mathbf{k}=%
\begin{cases}
\frac{2\pi}{Lp}\ (n_{1}\hat{\mathbf{e}}_{x_{1}}+n_{2}\hat{\mathbf{e}}_{x_{2}%
}),\\
\text{ with }n_{1},\ n_{2}\in\mathbb{Z}\ \\
\text{and }-\frac{L}{2}\leq n_{1}\pm n_{2}<\frac{L}{2}.
\end{cases}
\]

Therefore, we are demanding less crystal symmetry on the single particle HF
states to be obtained, since a lower number of constraints are being imposed
on the space of those states in which the solutions are sought.

\label{TB}

The Bloch basis which will be used for expanding the HF orbitals is written in
the form
\begin{align}
\varphi_{\mathbf{k}}^{(r,\sigma_{z})}(\mathbf{x},s)  & =\sqrt{\frac{2}{N}%
}\ u^{\sigma_{z}}(s)\sum_{\mathbf{R}^{(r)}}\exp(i\ \mathbf{k}\cdot
\mathbf{R}^{(r)})\ \varphi_{\mathbf{R}^{(r)}}(\mathbf{x}%
),\label{base de Bloch}\\
\hat{\sigma}_{z}u^{\sigma_{z}}  & =\sigma_{z}\ u^{\sigma_{z}},\nonumber\\
\varphi_{\mathbf{R}^{(r)}}(\mathbf{x})  & =\frac{1}{\sqrt{\pi a^{2}}}%
\exp(-\frac{(\mathbf{x}-\mathbf{R}^{(r)})^{2}}{2\ a^{2}}),\ a\ll p,\nonumber
\end{align}
where $N$ is the number of electrons in the system, $\hat{\sigma}_{z}$ is the
spin z projection operator, where $z$ is the orthogonal direction to the 2D
lattice; $\sigma_{z}=$ -1, 1, are the eigenvalue of the previously mentioned
operator and $\emph{r}=1,2,$ is the label which indicates each of the
sublattices. We are going to work on a half filling condition, then $N$
coincides with the number of cells in the square crystal with fixed periodic
boundary conditions $N_{c}$. Note that due to the tiny overlapping among
nearest neighbors approximation, the exact orthogonal character is only weakly
lost between elements corresponding to different sublattices and having the
same spin quantization. This fact is taken into account in the procedure for
solving the HF equations. That happens because some nearest neighbors belong
to different arrays. However,  the orthogonality between different elements
corresponding to the same sublattice, as well as unity norm for every
elements, is rigorously maintained. This follows because they are constructed
as Bloch states in their corresponding sublattices. \ Note that accordingly
with the TB approximation, it will be considered that the effective potential
created by the environment on each electron of the half filled band is a
quadratic function having a minimum on the lattice points and strongly
confining the electrons to it.

Now, let us consider the expansion of the HF orbitals in terms of the just
defined Bloch basis. The mean field single particle states incorporating the
before mentioned explicitly nonseparable form of the spin projections are given
as
\begin{equation}
\phi_{\mathbf{k},\ l}(\mathbf{x},s)=\sum_{r,\sigma_{z}}B_{r,\sigma_{z}%
}^{\mathbf{k},\ l}\varphi_{\mathbf{k}}^{(r,\sigma_{z})}(\mathbf{x}%
,s),\label{representaciontb}%
\end{equation}
where \emph{l} is the additional quantum number needed for indexing the
stationary state in question, which we are going to define precisely below.
After substituting (\ref{representaciontb}), (\ref{freehamiltonian}) and
(\ref{coulombio}) in (\ref{HF}); followed by projecting the obtained result on
the basis $\varphi_{\mathbf{k}^{\prime}}^{(\mathbf{t},\alpha_{z})}$ and some
extensive algebra, it is possible to arrive at the following self-consistent
matrix problem for the coefficients appearing in the expansion
(\ref{representaciontb}):
\begin{equation}
\lbrack\text{E}_{\mathbf{k}}^{0}+\widetilde{\chi}\ (\text{G}_{\mathbf{k}%
}^{dir}-\text{G}_{\mathbf{k}}^{ind}-\text{F}_{\mathbf{k}})]\mathbf{.}%
\text{B}^{\mathbf{k},l}=\widetilde{\varepsilon}_{l}(\mathbf{k})\ \text{I}%
_{\mathbf{k}}\mathbf{.}\text{B}^{\mathbf{k},l},\label{EcuMatricial}%
\end{equation}
where each of the quantities
\[
\text{B}^{\mathbf{k},l}=\bigl\|B_{(r,\sigma_{z})}^{\mathbf{k},\ l}\bigr\|,
\]
represents a vector having four components defined by the four possible pairs
$(r,\sigma_{z})$. The two appearing constants
\begin{align}
\widetilde{\chi}  & \equiv\frac{me^{2}}{4\pi\hbar^{2}\epsilon\epsilon_{0}%
}\frac{a^{2}}{p},\nonumber\label{e:coupling}\\
\widetilde{\varepsilon}_{l}(\mathbf{k})  & \equiv\frac{ma^{2}}{\hbar^{2}%
}\varepsilon_{l}(\mathbf{k}),
\end{align}
are dimensionless. In them, \emph{e} represents the charge of the particles;
$\hbar$ is the reduced Planck constant; \emph{a} is the characteristic radius
of the Wannier orbitals $\varphi_{0}$ while \emph{p} is the nearest neighbors
separation in the square lattice. It is clear now that we can define \emph{l}
= 1, 2, 3, 4, as a label indicating each of the four solutions to be obtained
for every value of quasi-momentum $\mathbf{k}$. Also, all the implicit
parameters in the 4$\times$4 matrices defining the HF problem
\begin{align}
\text{E}_{\mathbf{k}}^{0}  & =\bigl\|E_{\mathbf{k},(t,\alpha_{z}%
);(r,\sigma_{z})}^{0}\bigr\|\ ,\nonumber\\
\text{G}_{\mathbf{k}}^{dir}  & =\bigl\|G_{\mathbf{k},(t,\alpha_{z}%
);(r,\sigma_{z})}^{dir}\bigr\|\ ,\nonumber\\
\text{G}_{\mathbf{k}}^{ind}  & =\bigl\|G_{\mathbf{k},(t,\alpha_{z}%
);(r,\sigma_{z})}^{ind}\bigr\|\ ,\nonumber\\
\text{F}_{\mathbf{k}}  & =\bigl\|F_{\mathbf{k},(t,\alpha_{z});(r,\sigma_{z}%
)}\bigr\|\ ,\nonumber\\
\text{I}_{\mathbf{k}}  & =\bigl\|I_{\mathbf{k},(t,\alpha_{z});(r,\sigma_{z}%
)}\bigr\|\ ,\label{eq18}%
\end{align}
are dimensionless. The set of quantities (\ref{eq18}) constitute the matrix representations of
the periodic potential created by the mean field W$_{\gamma}$, the direct and
exchange terms in (\ref{HF}), the interaction potential with the neutralizing
\textquotedblright jellium\textquotedblright\ of charges F$_{b}$ defined in
(\ref{freehamiltonian}) and the overlapping matrix among the wavefunctions of the basis,
respectively. Each one of the four pairs, $(t,\alpha_{z})$ and $(r,\sigma
_{z})$ defines a row and a column respectively, of the matrix in question for
each momenta value. The explicit forms of the matrix elements are
given in Ref. \cite{symmetry} after assuming no restriction in the parameters. The normalization condition
 for the HF single particle states and the HF energy of the system take the forms%

\begin{align}
1  & =\text{B}^{\mathbf{k},l^{\ast}}.\text{I}_{\mathbf{k}}.\text{B}%
^{\mathbf{k},l},\\
E^{HF}  & =\sum_{\mathbf{k},l}\Theta(\widetilde{\varepsilon
}_{F}-\widetilde{\varepsilon}_{l}(\mathbf{k}))[\widetilde{\varepsilon}%
_{l}(\mathbf{k})-\frac{\widetilde{\chi}}{2}\text{B}^{\mathbf{k},l^{\ast}%
}.(\text{G}_{\mathbf{k}}^{dir}-\text{G}_{\mathbf{k}}^{ind}).\text{B}%
^{\mathbf{k},l}],
\end{align}
where $\Theta$ is the Heaviside function. The system (\ref{EcuMatricial}) is
non linear in the variables $B_{r,\sigma_{z}}^{\mathbf{k},\ l}$, which are the
four components of each vector $B^{\mathbf{k},l}$. Thinking in terms of
numerically solving the equations by the method of iterations, it is
convenient to pre-multiply them by $I_{\mathbf{k}}$ for each $\mathbf{k}.$
Note that for each $\mathbf{k}$ four eigenvalues (\emph{l} =$1,2,3,4.$) will
be obtained, or equivalently, four bands on the B.Z. From Eq. (\ref{EcuMatricial}),
it can be observed that in the representation
(\ref{base de Bloch}), the HF potentials and in general the total hamiltonian
of the system, resulted as being block diagonal with respect to the sets of
states indexed by the same \textbf{k}. This fact is a direct consequence of
the translation symmetries leaving invariant the sublattices.

In next section we will consider a half filling condition, that is, the HF solution will have one
electron per cell. In addition a periodic lattice of $N=20\times20$ electrons
will be also assumed. The occupied states inside the B.Z. are indicated in
figure \ref{f:ZB} a) by the points inside the shadowed region.

\section{SCES properties and overlapping controlled gap in the 2D TB models of
interacting particles}

In this section we present the \ results for the energy bands of the
\ considered TB model as determined by the solution of the system of equations
Eq. (\ref{EcuMatricial}) for the same system investigated  in
\cite{pla,symmetry,last} but assuming that the overlapping in the TB model is
changing. The change is implemented by modifying the parameter
$\ \ \widetilde{a}$. \ As in \cite{symmetry}, the solutions were found by
using the method of successive iterations. The parameter defining the width of
the jellium charge $\widetilde{b}=0.05$ is fixed. On the other hand, the
parameter $\widetilde{a}$ $=\frac{a}{p}$ is modified in order to
correspondingly vary the overlapping. The expression for $\widetilde{\gamma} $
(the quantity which defines the width of the free TB band associated with the
model as $4\widetilde{\gamma}$) is defined by the model in terms of
$\widetilde{a}$ as (See \cite{pla,symmetry,last})
\[
\widetilde{\gamma}=\exp(-\frac{1}{4\widetilde{a}})-0.05,
\]

\begin{figure}[h]
\includegraphics[width=6.0in,height=3.0in]{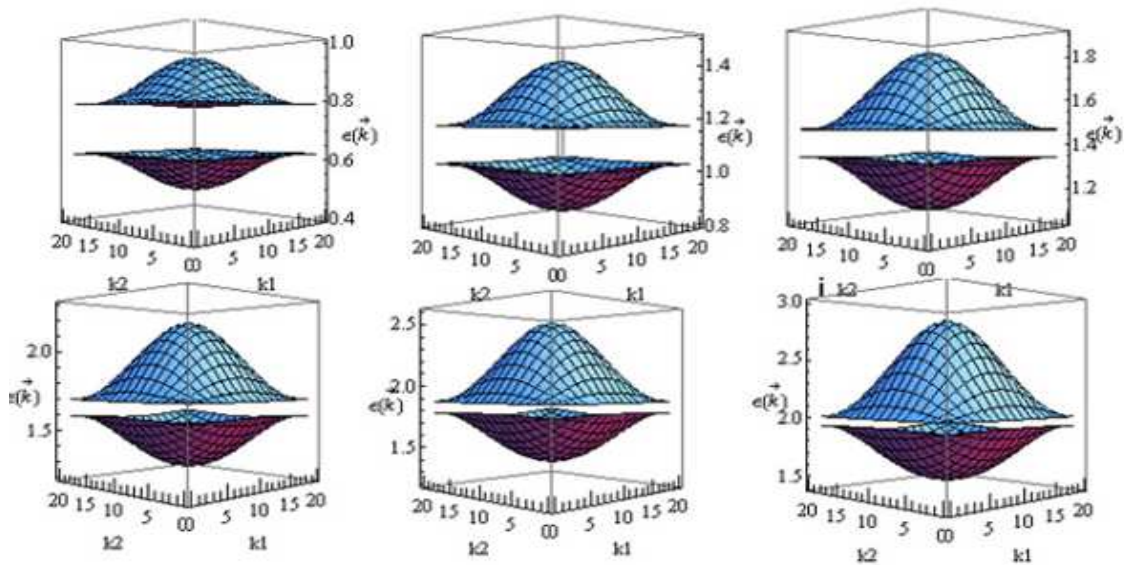}
\caption{The set of plots shows the band structure of the HF self-energy modes
for six values of the parameter $\hat{a}$ in the range $0.25-0.30$ in steps of
0.01. The value of $\hat{a}$ defines  the width (in units of $p$) of the Gaussian
 orbitals in the neighborhood of each lattice points. Note that
increasing overlapping by taking $\hat{a}$ larger, with the lattice fixed,
reduces the energy gap. }%
\label{f:bands}%
\end{figure}

\ The lattice unit cell was also fixed to the distance between the nearest
neighbor Cu atoms: $p=3.8$ \ A$^{0}$. The iteration process for solving the
equations started from a state selected to show antiferromagnetic order. This
allowed to achieve the convergence toward the solution presented in this
subsection, because there could exist a kind of barrier for the convergence
from an initial paramagnetic state. Figure \ref{f:bands}  shows a set
of band structures corresponding to varying values of the width of the
gaussian Wannier orbital. They correspond to lattices of 20x20 cells. The bands are depicted on the
scale of energies $ \frac {m a^2}{\hbar^2}$. \ The plots show how the insulator energy gap decreases
 as the overlapping increases when  the orbital size $\widetilde{a}$ becomes larger.
The results for the bands are almost in the thermodynamical limit for the
$20\times20$ lattice size employed, since the bands are practically unchanged
when a lattice of $30\times30$ unit cells is considered.  The presented calculation exemplifies
how a highly  unconstrained HF solution of a 2D model of TB interacting electrons is able to predict
 SCES properties as Mott types of insulator gaps.

\section{RTS in water doped graphite and 2D QDots lattices?}

In this section a possibility will be discussed for attaining room temperature
superconductivity (RTS) when graphite is doped with water on its surface. At the end
 of the section, and suggested by the analysis
 of water doped graphite system, some remarks indicating the possibility of RTS
in 2D lattices of Quantum Dots will be also advanced.
 The possibility of RTS in graphite  doped with water had been suggested
by recent experiments reporting the detection of granular room temperature
superconductivity in graphite powder, which was previously subject to
temperature treatment when suspended in water \cite{RT}. The measured
hysteresis loops in samples made with the powder, signaled the presence of
superconductivity and allowed to estimate a critical temperature higher
than room one. On the other hand, recent $ab-initio$ calculations of
clusters of water molecules over graphene in \cite{antwerp,wehling}, had
indicated that water molecules tend to form an hexagonal monoatomic layer on
top of the graphene, which can act as an imposed periodic external potential
on it. The authors remark that the deposition of a single layer on the
graphene leads to 2D planar hexagonal crystal of water molecules, staying
slightly above the graphite's last graphene plane.

\begin{figure}[h]
\par
\includegraphics[width=3.0in,height=3.0in]{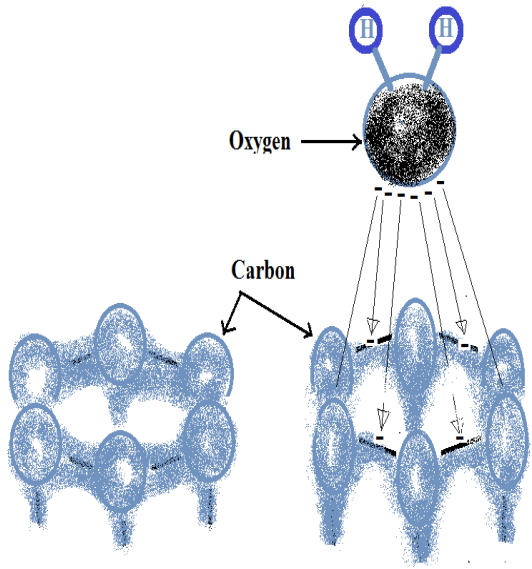}
\par
\caption{ The figure illustrates the possible action of the hexagonal array of
water molecules which $ab-inito$ calculations predict to be formed when water
is sufficiently doped with water molecules \cite{wehling,antwerp}. Only one
molecule of water is illustrated for simplicity. The hexagon of Carbon atoms
shown at the left, and not having the influence of the molecule of water, is
for comparatively illustrate the effect of the molecules on the electrons
clouds.The distances between the atoms are roughly proportion with the ones
defined by the simulations, and a last graphene layer in graphite was
considered, as indicated by its bounds with the next layer below. Note that
the electronegativity of Oxygen should tend to repel the electron density
along the C-C bonds. This effect can be also reinforced by the image of the
water dipole moment in graphite, which is a semi-metal. Thus, the possibility
exists that the action of the periodic potential created by a water monolayer
on graphite could generate a TB  band of the kind which  the cuprates show.
\cite{pla,symmetry,last}.}%
\label{sgraphite}%
\end{figure}

The centers of the Oxygen atoms rest directly over the centers of the
hexagons formed by the carbon atoms. They explain that the periodic potential
created by the water molecules on graphene does not affect it up to a point
in  which it becomes metallic. \ However, it was also argued that,  if a substrate
of $SiO_{2}$ is added below the graphene monolayer, then the system becomes
metallic. \ This is a property that we interpret as opening an opportunity for
obtaining RT superconductivity. The idea is described in what follows.
Firstly, graphite with a monolayer of water molecules over the last graphene
plane in it, can be considered as a graphene layer with a substrate of the
same graphite below it. \ But, as the simulations indicated, a substrate can
help the last layer to become metallic. The validity of this property is
directly supported by the fact that the same graphite is a semi-metal. Therefore,
assuming that the same predictions of SCES\ can be obtained from a TB model of
2D electrons with a lattice showing hexagonal structure, the last hexagonal
graphene layer on the surface of graphite (which is separated by a large
distance from the other layers) \ could eventually act as a TB metal, upon
which the deposited 2D crystal of water molecules exerts a strong dipolar
periodic electric potential. To illustrate this possibility consider
figure \ref{sgraphite} in which one of the water molecules doping the surface
of graphite is pictured according to the $ab-initio$ structure of a single
layer of water interacting with the surface of graphene.

\begin{figure}[h]
\includegraphics[width=4.0in,height=4.0in]{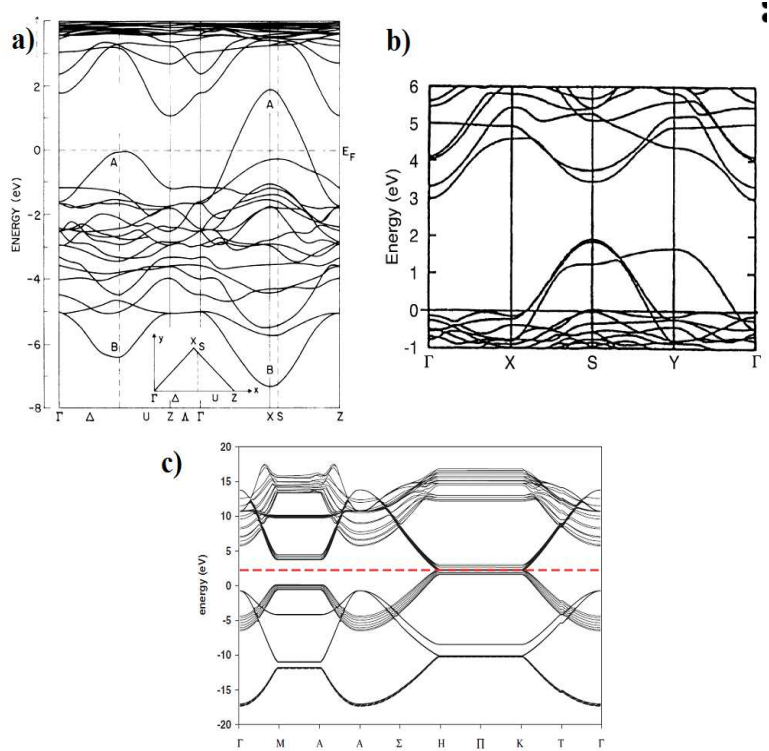}
\caption{The plot shows the band structures of three materials: a)
$La_{2}CuO_{4} $ (\cite{matheiss}), b) $YBa_2Cu_3O_7$ (\cite{YBCO}) and c)
Graphite (\cite{graphite}). Note that the bands crossing the Fermi level in
$La_{2}CuO_{4}$ have nearly 4 eV in energy width. On the other hand the same
kind of bands ranges only approximately 2 eV in energy in YBCO. Graphite bands being
closer from above and below to the Fermi level are very close in energy. }%
\label{3grafs}%
\end{figure}

Further, figure \ref{3grafs} shows the band
structures of $La_{2}CuO_{4}$ (\cite{matheiss}), $YBa_2Cu_3O_7$ (\cite{YBCO}) and
graphite (\cite{graphite}). Note from the plots that the bands crossing the
Fermi level in $La_{2}CuO_{4}$ have nearly $4$ eV in energy width. On the
other hand, the same type of bands range only approximately 2 eV in energy in
the $YBCO$ material. Then, let us consider that as the model of  $La_{2}CuO_{4}$
the width of the resulting TB like band given by the HF solution,
becomes roughly proportional to the resulting insulator gap,  after the constraints
on the HF scheme had been deleted. This assumption in turns indicates that in
the case of $YBCuO$, since the width of the TB like bands are nearly half
the ones of the $La_{2}CuO_{4}$, the insulator gap resulting of applying the
model to this material (let us call it $U$ in what follows) might be also nearly a half of the one obtained for
$La_{2}CuO_{4}$. But, note that the model considered here can be seen as kind of
improvement of the single-band Hubbard (or t-j) one. But, these models
indicate a critical temperature $T_c$ for superconductivity obeying  $T_{c}%
~\backsim\frac{t^{2}}{U}$. Therefore, assuming that the hopping parameter $t$
is similar for both cuprate materials, it follows that the prediction is in
agreement with the t-j model result, about that the $T_{c}$ of $YBCuO$
should nearly double the one of the $La_{2}CuO_{4}$. Therefore, after examining the band structure of
graphite, the following observation seems to support the possibility of
realizing higher critical temperatures in the graphite water systems. Note
from figure \ref{3grafs},  that the pure graphite bands which are closer from
above and below to the Fermi level, are very near in energy one to another.
This property suggests that the action of the hexagonal dipolar potential
created by the water molecules,  might be able to generate a TB binding band,
showing an energy width few times smaller than the  one in $YBCuO$
materials. Henceforth,  under the generation of the insulator HF
solution discussed here, this band might be able to determine a value of $U$
also few times smaller than the corresponding value in $YBCuO$.
In addition, the hopping parameter $t$ could be also possible to
become higher in the "graphite-water" system. Thus, the analyzed situation
suggests the possibility for the appearance of RTS in graphite doped with water as reported in \cite{RT}.
  Finally, it can be noted that the above discussion correctly  identify
that a small  insulator gap of the considered TB model could  contribute to
enhance $T_c$. This property opens the possibility of developing
2D nanotechnology systems being  described by a periodic TB models
of interacting electrons, which could enhance the nowadays attained critical temperatures.
 In particular, a  directly coming to the mind possibility is the growing of a 2D periodic array of Quantum Dots
 being driven  by an external potential. Then, by  varying the potential,
 the electron density can be to controlled, and a  hole doped 2D system being  similar to the CuO planes in $La_2CuO_4$ could be developed.  Therefore, assumed that not a  strong  experimental limitations arises
 for increasing the values of the parameter $\frac {t^2} {U}$ in the  considered 2D system, the possibility is suggested  of rising the  critical temperatures with respect to the ones attained in the cuprates. The study of the nowadays experimentally  realizable values for $\frac {t^2} {U}$ will be considered elsewhere.

\section*{Conclusions}

We consider former results for a Tight-Binding  model of CuO planes in
$La_{2}CuO_{4}$ with the objective of underlining their wider implications. It
is illustrated how interacting electrons systems described by the TB model can
show strongly correlated electron properties appearing in the mean field
approximation. For this outcome it becomes relevant to complement the HF
scheme by incorporating crystal symmetry breaking effects and non-collinear
spin orientations of the final HF orbitals. The results also furnish a general
theoretical insight for solving the old standing Mott-Slater debate, at least
for the $La_{2}CuO_{4}$ and TMO band structures. The idea suggested is to
apply a properly non-collinear GW scheme to the electronic structure
calculation of those materials. This view is strongly supported by the fact that
the GW method can be conceived as a generalized HF procedure, in which the
screening polarization is also determined. Thus, the possibility arises for
predicting the just phenomenologically fixed value for the dielectric constant
in Refs. \cite{pla,symmetry,last}, which is central in deriving the SCES
properties of $La_{2}CuO_{4}$ from the TB model in that references. Henceforth,
the whole discussion points to identify the main correlation properties in
these materials as mainly determined by screening. Preliminary arguments are
also presented indicating the possibility of attaining room temperature
superconductivity by means of a  doping with water molecules surfaces of
graphite, being orthogonal to its c-axis. The discussion opens an opportunity
to understand the recent reports of granular superconductivity at room
temperatures in graphite powder \cite{RT}. It should be mentioned that whole discussion
seem to be of help in the description of the observations of metal-insulator
transitions and Mott properties in systems of atoms trapped in planar photon
lattices \cite{opt1,opt2,opt3}. \ Finally, the analysis also furnishes  insights
about how to approach the old, but yet conflicting results about the appearance of
a magnetic order in the jellium model and the metallic systems at variable
electron densities \cite{march1, march2}.

\begin{acknowledgments}
N. H. M. and A. C. M. would like to acknowledge the kind hospitality of the
ASICTP during the Guest Scientists invitation to stay in the Centre during
August 2013. A. C. M. and A. C. B. also are grateful by the support received
from the Caribbean Network on Quantum Mechanics, Particles and Fields (Net-35)
of the ICTP Office of External Activities (OEA), the "Proyecto Nacional de
Ciencias B\'{a}sicas" (PNCB) of CITMA, Cuba.
\end{acknowledgments}


\begin{thebibliography}{99}
\bibitem {mott1}N.~F. Mott, \newblock \emph{Proc. Phys. Soc., London},
A62, 416, 1949.

\bibitem {slater1}J.~C. Slater, \newblock \emph{Phys. Rev.}
 82, 538, 1951.

\bibitem {slater}J.~C. Slater, \newblock \emph{Quantum Theory of Atomic
Structure}, Vol.~2, \newblock Dover Publications Inc., Mineola, New York, 1960.

\bibitem {dagoto}E.~Dagotto, \newblock \emph{Rev. Mod. Phys.} 66, 763, 1994.

\bibitem {bmuller}J.~G.~Bednorz and ~K.~A.~M\"{u}ller, \newblock \emph{Rev. Mod.
Phys.} 60, 585, 1987.

\bibitem {yanase}Y.~Yanase, \newblock \emph{Physics Reports} 387, 1, 2003.

\bibitem {almasan}C.~Almasan and ~M.~B.~Maple, \newblock \emph{Chemistry of High
Temperature Superconductors}, \newblock World Scientific, Singapore, 1991.

\bibitem {pickett}W.~E. Pickett, \newblock \emph{Rev. Mod. Phys.} 61, 433, 1989.

\bibitem {imada}M.~Imada, \newblock \emph{Rev. Mod. Phys.} 70, 4, 1998.

\bibitem {anderson1}P.~W. Anderson, \newblock \emph{Phys. Rev.} 115, 2, 1959.

\bibitem {hubbard}F.~Hubbard, \newblock \emph{Proc. Roy. Soc., London},
A276, 238, 1963.

\bibitem {gutzwiller}M.~Gutzwiller, \newblock \emph{Phys. Rev.} A134, 923, 1964.

\bibitem {gutzwiller1}M.~Gutzwiller, \newblock \emph{Phys. Rev.} A137, 1726, 1965.

\bibitem {terakura}A.~R. Williams, N.~Hamada, K.~Terakura and  T.~Oguchi,
\newblock \emph{Phys. Rev. Lett.} 52, 1830, 1984.

\bibitem {vanharlingen}D.~J.~Van Harlingen, \newblock \emph{Rev. Mod. Phys.}
67, 515, 1995.

\bibitem {damascelli}A.~Damascelli, Z. Hussain and Zhi-Xun Shen,
\newblock \emph{Rev. Mod. Phys.} 75, 473, 2003.

\bibitem {Burns}G.~Burns, \newblock \emph{High-Temperature Superconductivity},
\newblock Academic Press, New York.

\bibitem {freltoft}G.~Shirane, D. E. Moncton,  S. K. Sinha,  S. Vaknin, J. P.
Remeika, A. S. Cooper, D.~Harshman,  T.~J.~Freltoft and  J. E.~Fischer, \newblock
\emph{Phys. Rev.} B36, 826, 1987.

\bibitem {deBoer}J.~H. de~Boer~ and  E.~J. W.~Verway, \newblock \emph{Proc. Phys.
Soc., London}, A49, 94, 1937.

\bibitem {peierls}R.~Peierls, \newblock \emph{Proc. Roy. Soc., London}
A49,72, 1937.

\bibitem {adersson}P.~W. Anderson, \newblock \emph{Science} 235, 1196, 1987.

\bibitem {fradkin}E.~Fradkin,  \newblock \emph{Field Theories of Condensed
Matter}, Vol.~82. \newblock Addison Wesley Publishing Company, 1991.

\bibitem {rice}W.~F.~Brinkman and ~T.~M.~Rice, \newblock \emph{Phys. Rev.}
B2, 4302, 1970.

\bibitem {kohn}W.~Kohn, \newblock \emph{Phys. Rev.} A133, 171, 1964.

\bibitem {kohn1}W.~Kohn and ~L.~J.~Sham, \newblock \emph{Phys. Rev.} A140, 1133, 1965.

\bibitem {singh}D.~J.~Singh and ~W.~E.~Pickett, \newblock \emph{Phys. Rev.}
B44, 7715, 1991.

\bibitem {szabo}A.~Szabo and ~N.~Ostlund, \newblock \emph{Modern Quantum
Chemistry: Introduction to Advanced Electronic Structure Theory},
\newblock Dover Publications Inc., Mineola, New York, 1989.

\bibitem {fetter}A.~L.~Fetter and ~J.~D.~Walecka, \newblock \emph{Quantum Theory
of Many Particle Physics}, Vol.~1, \newblock McGraw-Hill, Inc., 1971.

\bibitem {matheiss}L.~F. Matheiss, \newblock \emph{Phys. Rev. Lett.} 58, 1028, 1987.

\bibitem {mott}N.~F. Mott, \newblock \emph{Metal-Insulator Transition}.
\newblock Taylor and Francis, London/Philadelphia, 1990

\bibitem {dirac}P.~A.~M. Dirac, \newblock \emph{Proc. Cambridge. Phil. Soc.}
26, 376, 1930.

\bibitem {gebauer}R. Gebauer, S. Serra, G. L. Chiarotti, S. Scandolo, S.
Baroni and E. Tosatti, \emph{Phys. Rev.} B61, 6145, 2000.

\bibitem {tesis}A. Mosca Conte,\textit{\ Quantum mechanical modeling of nano
magnetism, } Ph. Dissertation Thesis, Interantional School of Advanced
Studies, Trieste, Italy, February 2007.

\bibitem {augustoref}B. J. Powel,\textit{\ An Introduction to Effective
Low-Energy Hamiltonians in Condensed Matter Physics and Chemistry.}
arXiv:0906.1640v6 2009, physics.chem-ph.

\bibitem{march}G. Forte, G.G.N. Angilella, V. Pittalaa, N.H. March and R. Pucci,
\textit{\ Phys. Lett.} A 376, 476.479, 2012.

\bibitem{march1}F. Herman and N. H. March, \textit{Sol. State Comm. } 50, 25, 1984.

\bibitem{march2}W. Jones and N. H. March, \textit{Theoretical Solid State
Physics, Volume 1: Perfect Lattices in Equilibrium}, Physics $\&$ Astronomical
Monographs 27, John Wiley $\&$ Sons Ltd. 1973.

\bibitem{pla}A. Cabo-Bizet and A. Cabo, \textit{Phys. Lett.} A 373, 1865, 2009.

\bibitem{symmetry}A. Cabo-Bizet and A. Cabo, \textit{Symmetry} 2, 388, 2010.

\bibitem{last}V. M. Martinez Alvarez, A. Cabo-Bizet and A. Cabo,
\textit{Quantum phase transition from a Antiferromagnetic-Insulator to a
Paramagnetic-Metal laying beneath the superconducting dome},
 arXiv:1210.6548v2 [cond-mat.str-el], 2012.

\bibitem {su}Y-S. Su, T. A. Kaplan, S. D. Mahanti and J. F. Harrison,
\textit{Phys. Rev. } B 59, 10521, 1999.

\bibitem {perry}J. Perry, J. Tahir-Kheli, W. A. Goddard, \textit{Phys. Rev.} B
63, 144510, 2001.

\bibitem {RT}T. Scheike, W. Bohlmann, P. Esquinazi, J. Barzola-Quiquia, A.
Ballestar and A. Setzer, \textit{\ Adv. Mater.} 24, 5826, 2012.

\bibitem {laughlin}R. B. Laughlin, \textit{Hartree-Fock Computation of the
High-Tc Cuprate Phase Diagram}, arXiv:1306.5359v1 [cond-mat.supr-con], 22 Jun. 2013.

\bibitem {wehling}T. O. Wehling, A. I. Lichtenstein M. I. Katsnelson, \textit{App.
Phys. Lett.} 93, 202110, 2008.

\bibitem {antwerp}O. Leenaerts, \textit{An ab initio study of the adsorption
of atoms and molecules graphene}, PhD Degree Dissertation, Department of
Physics, University of Antwerp, Antwerp, Belgium, 2010.

\bibitem {opt1}W. Hofstetter, J. I. Cirac, P. Zoller, E. Demler and M. D.
Lukin, \textit{Phys. Rev. Lett.} 89, 220407-4, 2002.

\bibitem {opt2} K. Drese and M. Holthaus, \textit{Phys. Rev. Lett.} 78, 2932, 1997.

\bibitem {opt3} M. Greiner, O. Mandel, T. Esslinger, T. W. Hansch and I. Bloch,
\textit{Nature} 415, 39, 2002.

\bibitem {YBCO}  K. W. Wong and  W. Y. Ching, \textit{Physica C} 416, 47, 2004.

\bibitem {graphite}   N. Ooi, A. Rairkar and  J. B. Adams, \textit{Carbon} 44, 231–242, 2006.
\end{thebibliography}
\end{document}